# Quantum Zeno Effect Underpinning the Radical-Ion-Pair Mechanism of Avian Magnetoreception

Iannis K. Kominis[1,2]

[1]*Department of Physics, University of Crete, Heraklion 71103, Greece*

[2]*Institute of Electronic Structure and Laser, Foundation for Research and Technology,*

*Heraklion 71110, Greece*

**The intricate biochemical processes underlying avian magnetoreception, the sensory ability of migratory birds to navigate using earth's magnetic field, have been narrowed down to spin-dependent recombination of radical-ion pairs to be found in avian species' retinal proteins. The avian magnetic field detection is governed by the interplay between magnetic interactions of the radicals' unpaired electrons and the radicals' recombination dynamics. Critical to this mechanism is the long lifetime of the radical-pair's spin coherence, so that the weak geomagnetic field will have a chance to signal its presence. It is here shown that a fundamental quantum phenomenon, the quantum Zeno effect, is at the basis of the radical-ion-pair magnetoreception mechanism. The quantum Zeno effect naturally leads to long spin coherence lifetimes, without any constraints on the system's physical parameters, ensuring the robustness of this sensory mechanism. Basic experimental observations regarding avian magnetic sensitivity are seamlessly derived. These include the magnetic sensitivity functional window and the heading error of oriented bird ensembles, which so far evaded theoretical justification. The findings presented here could be highly relevant to similar mechanisms at work in photosynthetic reactions. They also trigger fundamental questions about the evolutionary mechanisms that enabled avian species to make optimal use of quantum measurement laws.**

Magnetic interactions in radical-ion pairs *(1)* hosted in photosensitive proteins are currently thought to be at the basis of avian magnetoreception, the biological compass that avian species (as well as amphibian and marine species *(2)* ) use for navigation. The retinal localization *(3,4)* of cryptochrome *(5)* proteins strongly supports this theory since it is the photoexcited donor-acceptor molecules found in cryptochromes that initiate the magnetoreception mechanism by the creation of the radical-ion pair. Further evidence from radiofrequency resonance experiments *(6)* corroborates the working model *(7)* of avian magnetic field detection. This is understood to arise from the interplay between magnetic interactions *(8,9)* of the radical-pair's unpaired electrons and the radical's recombination dynamics. Specifically, if the radical-ion-pair (RIP) is initially created in the singlet spin state (electron spins anti-aligned), hyperfine interactions with the molecule's nuclear spins and Zeeman interaction with earth's magnetic field start mixing some triplet (electron spins aligned) character into the system's quantum state. This mixing results in a temporal oscillation of the RIP's quantum state (not unlike a precession of a classical top), known as the singlet-triplet (S-T) quantum coherence. Since triplet-state pairs, if at all, recombine to different chemical products than singlet-state pairs, the magnetic field affecting the mixing is encoded in the change of e.g. the singlet-state product yield. Further processing transforms this information into a visual perception *(7)* that determines the avian response.

It is clear that the S-T coherence lifetime is central to the success of this mechanism. Any process interrupting coherent S-T mixing must be slow enough for appreciable mixing to occur, and measurable yield changes to follow suit. Like every natural oscillator, however, the S-T coherence is susceptible to damping. The RIP recombination process was until now thought to be such a damping mechanism, and the associated recombination rates were thus de-facto required to be smaller than mixing rates. At earth's magnetic field of 0.5 G the latter are on the order of 1 $\mu s^{-1}$. It is known, however, that typical recombination rates are *(10-12)* on the order of 10 μs$^{-1}$ or higher, and depend *(5)* exponentially on parameters such as the distance between the radical-pair's donor and acceptor molecules. How then can the RIP magnetoreception mechanism function?

In this article it is shown that quantum physics comes to the rescue of the magnetoreception mechanism's viability. One of the peculiarities of quantum physics is that the mere act of observing a quantum system alters its evolution. In particular, frequently observed dissipative quantum systems (damped quantum oscillators) decay at different rates than when freely evolving or weakly observed. This is the well-known quantum Zeno effect, theoretically introduced *(13)* in 1977. This effect has been observed in numerous experimental settings, ranging from optical transitions *(14)* of trapped ions to the collision-induced slowing down of ortho-para conversion in molecular spin isomers *(15)*.

As will be detailed in the following, the RIP's recombination process constitutes a quantum measurement performed on the pair's spin state. Naturally occurring high recombination rates lead to the appearance of the quantum Zeno effect, automatically securing long S-T coherence lifetimes. The counter-intuitive nature of the quantum Zeno effect is manifested in the fact that the higher the recombination rate, the longer the lifetime of S-T beating becomes. A strikingly similar effect underlies recent advances *(16)* in atomic magnetometers, the magnetic sensitivity of which critically depends on the relevant spin coherence lifetime. Based on the theory to be developed, all experimental findings regarding avian magnetic sensitivity are effortlessly derived in a self-consistent way, without even solving the relevant evolution equation for the RIP's spin state. Observations regarding the value of avian magnetic sensitivity, the birds' insensitivity to field-polarity, the heading error in bird orientation experiments and the magnetic sensitivity functional window, the latter two of which so far evaded theoretical justification, directly follow from the eigenvalue-spectrum of the RIP's spin state evolution equation. Before proceeding with the demonstration of the quantum-mechanical basis of avian magnetoreception, the workings of the well-established atomic magnetometers will be briefly dwelled upon, since these two seemingly disparate physical realizations of magnetometry share the same physical foundations.

**Quantum Zeno effect in atomic magnetometers**

The magnetic sensitivity of an ensemble of $N$ atoms with gyromagnetic ratio $\gamma$ is *(17)* $\delta B = 1/\gamma\sqrt{N\tau T}$, where $\tau$ and $T$ are the spin coherence lifetime and the total averaging time, respectively. Early atomic magnetometers *(18)* were dominated by spin-exchange collisions happening at a characteristic rate $1/\tau_{se}$ (proportional to atom density) and damping atomic spin coherence at the same rate. Low atom densities were thus unavoidable, leading to a relatively low magnetic sensitivity (or high $\delta B$). However, as realized by Happer *(19)*, the daunting effect of spin-exchange collisions can be suppressed, if the spin-exchange rate is much larger than the other frequency scale of the problem, the Larmor frequency $\omega$ of the spin precession in the applied magnetic field. Under this condition, $\omega\tau_{se} \ll 1$, it has been shown that the spin coherence lifetime is increased to $\tau' \sim \tau_{se}/(\omega\tau_{se})^2 \gg \tau_{se}$. The resulting dramatic improvement in sensitivity was enunciated by Romalis *(20,21)* in the development of a new atomic magnetometer, which reached a subfemtotesla magnetic sensitivity, and which infiltrated a wide range of scientific disciplines, ranging from biomagnetic imaging of the human brain *(22)* and low-field NMR *(23)* to precision gyroscopes *(24)* and tests of fundamental symmetries of nature *(25)*.

The suppression of spin-exchange relaxation in atomic vapors is yet another manifestation (apparently the first experimentally observed) of the quantum Zeno effect. Quantum Zeno effects in general involve two rates: the intrinsic oscillation frequency of a system and the interrogation rate of the system's state. If the latter exceeds the former, the effective decay rate of the system is suppressed. In dense atomic vapors used in optical magnetometers, spin-exchange collisions effectively probe the atomic spin precession. For two colliding atoms with electron spins $s_1$ and $s_2$, the spin-exchange interaction *(26)* damping spin coherence is of the form $s_1 \cdot s_2$. This interaction can be interpreted as a measurement performed by the atom 2 on atom 1 (and vice versa). If the rate of probing (collision rate) significantly exceeds the Larmor frequency, the quantum Zeno effect surfaces, leading to a counter-intuitive rise of the spin coherence lifetime *(27)*.

**Radical-ion-pair recombination is a quantum measurement**

The recombination process of radical-ion pairs is a quantum measurement *(28, 29)* which effectively probes the observable $Q_S = 1/4 - s_1 \cdot s_2$, the singlet-state projection operator. In other words the recombination process itself is interrogating the RIP whether it is in the singlet or in the triplet state, so that it can recombine accordingly. This interrogation is a damping mechanism for the S-T coherence, much as atomic spin-exchange collisions dissipate atomic spin coherence. However, when the interrogation frequency is high enough relative to the frequency scale set by magnetic interactions in the RIP, the decay of the S-T coherence is suppressed. The result of the measurement of $Q_S$ is either 1 (meaning the radical-pair is in the singlet state), or 0 (meaning the pair is in the triplet state). After one or the other alternative is realized, chemical reactions transform the radical-pair to its state-specific products. The above processes are depicted schematically in Fig. 1.

The time evolution of the RIP's spin state is described by a differential equation (Liouville equation) involving the pair's density matrix, which contains all information on the RIP's quantum state. Most if not all of the system's basic physical properties follow from the eigenvalue spectrum of the Liouville equation. Specifically, the time evolution of the RIP's spin state can be described by a discrete number of modes each one characterized by a decay rate and a precession frequency. These are encoded in the complex eigenvalues of the Liouville equation, which are of the form $-\lambda + i\Omega$, with $\lambda \geq 0$ being the decay rate and $\Omega$ the precession frequency of the S-T coherence, respectively. The number of these modes is $N_d^2$, where $N_d$ is the dimension of the density matrix determined by the number of interacting spins in the RIP donor and acceptor molecules (see Methods summary). A simple four spin-1/2 RIP (two unpaired electrons and two nuclear spins) is considered here as it is enough to illustrate the essential physical arguments. In Fig. 2a the decay rates of the eigenvalue spectrum are displayed. The emergence of the quantum Zeno effect is readily observed in the regime where the recombination rate $k_s$ is much larger than the Larmor frequency $\omega$ (at earth's field of 0.5 G, $\omega \approx 0.7 \ \mu s^{-1}$). When $k_s / \omega \gg 1$, the decay rates split into two branches. The "ordinary" decay rates (upper branch) which exhibit the expected proportional scaling with the rate $k_s$, and a lower branch in which the decay rates decrease with increasing $k_s$. The latter manifest the counter-intuitive nature of the quantum Zeno effect: in the presence of a strong dissipative mechanism (large $k_s$), the S-T quantum coherence survives for a longer time.

**Explanation of experimental observations on the avian magnetic compass**

Just by counting eigenvalues it is straightforward to arrive at an explanation of all basic experimental findings. This is because the short-lived modes (upper branch in Fig. 1a) will decay away in a time scale too short for earth's magnetic field to have any effect. On the other hand, the long-lived modes will support coherent S-T mixing for a time long enough to allow the geomagnetic field to materialize its presence by altering chemical product yields. The larger the number of long-lived eigenvalues, the stronger will be the S-T mixing, and the more pronounced the change in chemical product yields.

The lower branch of Fig. 2a is found to contain about 60% of the total number of eigenvalues. These are responsible for long-lived terms contributing to the S-T coherence. Each one of those terms will survive for a time on the order of $\tau = 1/\lambda >> \omega$, and will sweep a phase on the order of $\Omega\tau >> 1$, allowing many mixing cycles to occur. On average, half of those terms will be found to be in the triplet state upon the recombination-induced interrogation and will lead to triplet-state products. The singlet-product yield will thus be reduced by roughly 30%, in very good agreement with previous estimates *(30)*. More important, this happens irrespective of the specific value of the rate $k_s$ and irrespective of the number of nuclear spins (i.e. the number of eigenvalues). Thus the magnetoreception mechanism is quite robust, being insensitive to specific values of the system's parameters, as should be the case for the sensory mechanism of living species. In Fig. 2(b) the rate $k_s$ is kept constant while varying $a/\omega$, the ratio of the hyperfine to Larmor frequency. The observed crossing of upper-branch (short-lived) eigenvalues into the lower-branch (and vice versa) underlies the magnetic sensitivity effect. Indeed, Fig. 2(c) shows the percentage decrease, $Y_d$, of the singlet-product yield. This is calculated based on the previous argument, i.e. if the number of long-lived decay rates (those for which $\lambda < \omega$) is $N$, then the singlet-product yield is expected to decrease by $N/2N_{tot}$, where $N_{tot}$ is the total number of eigenvalues (256 in this case). As is evident from Fig. 2(c), for a specific value of the hyperfine frequency $a$, a window $\Delta\omega$ of Larmor frequencies (or magnetic fields) exists where there is a non-zero change in $Y_d$. If this window is centered at the earth's field of 0.5 G (this happens in this case if $a=8\omega=4$ G, a reasonable hyperfine coupling), it follows by inspection that the slope of the changing part of the plot is $\Delta Y_d / \Delta\omega \approx 15 \% / G$, implying that the observed avian sensitivity of 0.01 G requires a sensitivity of measuring product-yield change at the level of 0.2%, in excellent agreement with existing estimates *(31)*. Whereas current theoretical models *(9)* predict a continuous avian response starting from zero magnetic field, this plot explains in a straightforward way the experimentally observed sensitivity window *(7,32)*. It is stressed that the appearance of the sensitivity window is due to the finite number of eigenvalues and the crossing of the two branches of eigenvalues seen to occur by varying $a/\omega$, i.e. for too small or too large values of the ratio $a/\omega$ no crossing occurs. The above results are valid for any value of the recombination rate $k_s$, as long as there is a large enough hyperfine coupling. For example if $k_s = 100\omega \approx 0.1\ \text{ns}^{-1}$ the hyperfine coupling must be $a = 20\omega \approx 10\ \text{G}$ to create the same sensitivity window.

The magnetic-field position of the center of the sensitivity window obviously depends on the hyperfine frequency $a$. Indulging in any evolution-based interpretations is by no means the current work's objective. Based on the previous analysis, however, it seems that the only requirement for magnetoreception to work is for the hyperfine frequency $a$ to have a value such that the sensitivity window is properly centered at the geomagnetic field. Thus the radical-pair's molecular structure, and in particular the nuclear species determining the unpaired electrons' hyperfine couplings must have evolved in such a way as to properly tune the hyperfine frequencies. This is a more plausible scenario than assuming that nature had conjured up a parameter-sensitive magnetoreception mechanism and then evolved the parameters (recombination rates) until everything works out. After all, typical bio-molecular hyperfine couplings are *(33)* between 1 and 10 G, so proper centering should not have been too difficult to realize.

In Fig. 3 the angular dependence of magnetoreception is addressed. Orientation experiments with bird ensembles *(34,35)* have observed that the avian magnetic compass is not perfect, i.e. there is a heading angle error. The heading angle, defined on the horizon plane, is the angle between the bird's flight direction and the magnetic field. All such observations report heading angle distributions with a standard deviation of about 15°. This heading error is derived in the following. In Fig.3a the lower-branch (long-lived) decay rates are shown as a function of the heading angle $\varphi$. It is noted that every long-lived mode of the S-T coherence will have a simple time dependence of the form $e^{-(\lambda_i + i\Omega_i)t}$ and will survive for a time on the order of $1/\lambda_i$, roughly contributing a term $y_i \sim \cos(\Omega_i / \lambda_i)$ to the singlet-state product yield. In Fig.3b the phases $\Omega_i / \lambda_i$ are shown as a function of the heading angle $\varphi$. The magnetic-sensitivity dependence on heading is attributed to the modulation of these phases by the heading angle. In Fig.3c the average of $y_i$ over all long-lived eigenvalues is plotted as a function of $\varphi$. The first observation is the reflection symmetry about $180^{\circ}$, proving the insensitivity of the magnetoreception mechanism to magnetic field polarity. Second, the signal height is $S \approx 0.25$ whereas the "noise" level, stemming from the superposition of several oscillatory terms, is $N \approx 0.05$. The full width at half maximum of each of the two broad dips is $\Delta\varphi \approx 90^{\circ}$ leading to a heading angle sensitivity (heading error)

$$\delta\varphi \approx \frac{\Delta\varphi}{S / N} \approx 18^{\circ}, \tag{1}$$

in very good agreement with the measured values. It noted that the avian heading error is attributed to the well-known Gibbs phenomenon of Fourier analysis, i.e. the wiggles appearing in the angular response are due to the superposition of a finite number of eigenvalues. By adding one more nuclear spin in the system, raising the number of eigevalues from 256 to 1024, the "noise" level is seen to persist.

In summary, it is rather captivating a realization that the eigenvalue spectrum of a density matrix determines macroscopic behaviour of living species. Even more so is the fact that a quantum sensor is at work in magnetic-sensitive birds, i.e. a fundamental quantum phenomenon so far understood to affect truly microscopic and well-isolated quantum systems is seen to underpin a sensory mechanism of avian species and their associated biological behaviour. This realization could lead to the discovery of several other radical-ion pairs participating in similar mechanisms in avian or other species, which so far were not considered due to apparently high recombination rates.

Concluding, the following fundamental questions naturally follow from the presented work: (a) what is the relation of the reported findings to similar effects involved in photosynthetic *(36,37)* reactions? (b) Is the ability of magnetic-sensitive species to make optimal use of basic quantum measurement principles coincidental or the result of an evolutionary mechanism? (c) Finally, notwithstanding the more philosophical nature of this question, is there any relation of the exponentially rich eigenvalue spectrum of magnetically-active bio-molecules to increasingly complex biological behaviour?

## Methods

Since avian species processing sensory signals at the bio-molecular level lack the ability of averaging, the measurement time of their magnetoreceptor is determined by the decay time $\tau$ of the S-T quantum coherence. Hence, from Heisenberg's energy-time uncertainty, the single-molecule magnetic sensitivity cannot be larger than $1/\gamma\tau$, where $\gamma$ is the unpaired electron gyromagnetic ratio (for a free electron, $\gamma$=2.8 MHz/G). If the number of radical-pairs participating in the measurement is $N$, then the signal-to-noise ratio is enhanced by $\sqrt{N}$, increasing the magnetic sensitivity to $\delta B = 1/\gamma\tau\sqrt{N}$. This is a fundamental limit independent of the particular realization of the measurement process, which might not be able to reach this sensitivity limit. In avian mangetoreception it is the minimum detectable change in the chemical product yields that actually limits magnetic sensitivity. If the S-T coherence oscillates at frequency $\Omega \approx \gamma B$ and survives for a time $\tau$, then the drop in the singlet-product yield will roughly be $y_d \approx 1 - \cos(\omega\tau) \approx (\gamma B\tau)^2$. Since the minimum detectable yield change is *(31)* $\delta y \approx 0.1\%$, it follows that the minimum detectable magnetic field (or field change) is

$$\delta B = \frac{\sqrt{\delta y}}{\gamma\tau} \qquad (2)$$

If the number of retinal radical-ion pairs is *(31)* $N \approx 10^8$, the actual magnetic sensitivity is seen to fall short of the fundamental limit by about two orders of magnitude. For an S-T coherence lifetime $\tau \leq 0.01 \, \mu s$, it follows from (2) that $\delta B \approx 1 \, G$, clearly inadequate to detect earth's magnetic field, let alone small variations thereof. Thus the lifetime $\tau$ must be larger than 1 µs in order to explain observed avian magnetic sensitivities reaching *(38)* 1 mG. The quantum Zeno effect guarantees that such a long lifetime of the S-T quantum coherence will exist.

To elucidate the quantum Zeno effect a simple spin-1/2 system in a magnetic field $B$ is considered first. Frequent measurements of the system observable $s_x$, the x-axis spin projection, are performed at a rate $1/\tau$. It is readily shown following standard methods of quantum measurement theory *(28,29)* , that the system's time evolution is described in terms of its density matrix by the Liouville equation ( $\hbar = 1$ )

$$\frac{d\rho}{dt} = -i[H, \rho] - \frac{1}{\tau}[s_x, [s_x, \rho]] \tag{3}$$

where $H = \omega s_z$ is the Zeeman interaction with the applied magnetic field. There are two cases involving the problem's two frequency scales: if $\omega\tau \gg 1$ it follows that the coherence $\langle s_\perp \rangle = \langle s_x \rangle + i \langle s_y \rangle$ precesses about the magnetic field at frequency $\omega$ and decays exponentially with time constant $2\tau$, which is the case usually considered and intuitively understood. However, if $\omega\tau \ll 1$, it follows that the coherence decays with two time constants, $\tau_1 = \tau$ and $\tau_2 = \tau / 2\omega^2\tau^2 \gg \tau$, the latter governing its long-time evolution.

In the case of the radical-ion-pairs, the observable probed by the recombination process is the singlet-state projection operator $Q_S$. The time evolution of the radical-pair's density matrix is given by equation similar to (2)

$$\frac{d\rho}{dt} = -i[H,\rho] - k_s[Q_S,[Q_S,\rho]] \tag{4}$$

where the first term describes the magnetic interactions in the RIP and the second the continuously performed quantum recombination measurement. Only the singlet recombination term is kept, since similar conclusions follow if we also add the triplet recombination term $-k_t[Q_T,[Q_T,\rho]]$. Since in most cases *(12)* $k_t \ll k_s$, the triplet term is omitted altogether. The probability $S$ ($T$) to find the radical-pair in the singlet (triplet) spin state is $S = \mathrm{Tr}\{\rho Q_S\}$ ($T = \mathrm{Tr}\{\rho Q_T\}$). While the system is coherently beating at the mixing frequency $\Omega$, keeping $S+T=1$, it will perform a quantum jump *(39)* to either the singlet ($S=1$) or the triplet state ($T=1$) leading to corresponding chemical products. These jumps will happen at a rate defined by several time constants, given in terms of the real parts of the calculated eigenvalues. If equation (4) is written as $d\rho/dt = L(\rho)$, where $L(\cdot)$ is the Liouville super-operator, diagonilizing the matrix $A$ for which $L(\rho) = A\rho$ ( $\rho$ is now a column vector containing all density matrix elements) leads to the eigevalues used in all calculations. The time evolution equation (4) is derived from fundamental quantum measurement theory. The reason that previous works had to invoke unrealistically small recombination rates is that in the phenomenological evolution equations used *(12)*, the recombination reaction kinetics and the quantum evolution of the radical-pair's spin state were intermingled in a single density matrix evolution equation. The S-T coherence decay rate was thus directly proportional to the recombination rates, which had to be kept small in order for the weak magnetic field to be able to measurably affect product yields. An attempt to reconcile the problem has appeared early on in the literature *(40)*.

For almost all calculations a simple model for a RIP was considered, consisting of two molecules (donor and acceptor) having one spin-1/2 nucleus each, denoted by $\boldsymbol{I}_1$ and $\boldsymbol{I}_2$. The Hamiltonian is $H = H_1 + H_2$, with $H_i = \omega\hat{\boldsymbol{b}}\cdot\boldsymbol{s}_i + a\boldsymbol{I}_i\cdot\ddot{g}_i\cdot\boldsymbol{s}_i$ ($i$=1,2) being the Zeeman and hyperfine interaction energy of each electron with the applied magnetic field $\boldsymbol{B} = B\hat{\boldsymbol{b}}$ (pointing along the unit vector $\hat{\boldsymbol{b}}$) and the molecule's only nucleus. The Larmor frequency, $\omega$, and the hyperfine frequency, $a$, were taken to be the same for both electrons. The diagonal hyperfine tensors $\vec{g}_1$ and $\vec{g}_2$ describe the anisotropy of the hyperfine interactions and fix the local coordinate system of the avian magnetoreceptor. Without loss of generality, the values $g_{1,xx} = g_{1,yy} = 1$, $g_{1,zz} = 0$ and $g_{2,xx} = g_{2,yy} = g_{2,zz} = 1$ were used. For the angular dependence calculations only, the value $g_{2,yy} = 2$ was used in order to break the symmetry in the x-y plane. Exchange and dipolar interactions *(41)* do not alter the relevant physics and were ignored. Spin relaxation effects can also be neglected, as the Rabi frequency producing disorientation in the resonance experiments *(6)* is on the order of 0.1 $\mu s^{-1}$, hence any spin relaxation effects must result in relaxation times larger than 10 $\mu s$, long enough not to have any adverse effect (after all, birds do orient). Further predictive power of the theory here developed will be elaborated upon in more detail elsewhere. For example, earlier observations *(42)* of unusually high triplet-product lifetimes observed with the cyclic dipeptide molecule cyclo(DkNAp-ThQx) can be explained.

**Acknowledgements** I wish to acknowledge Dr. D. Anglos for bringing to my attention his early experimental observations of unusually high radical-ion-pair triplet lifetimes, Prof. D. Charalambidis and Prof. P. Rakitzis for several helpful comments.

**Correspondence** should be addressed to ikominis@iesl.forth.gr

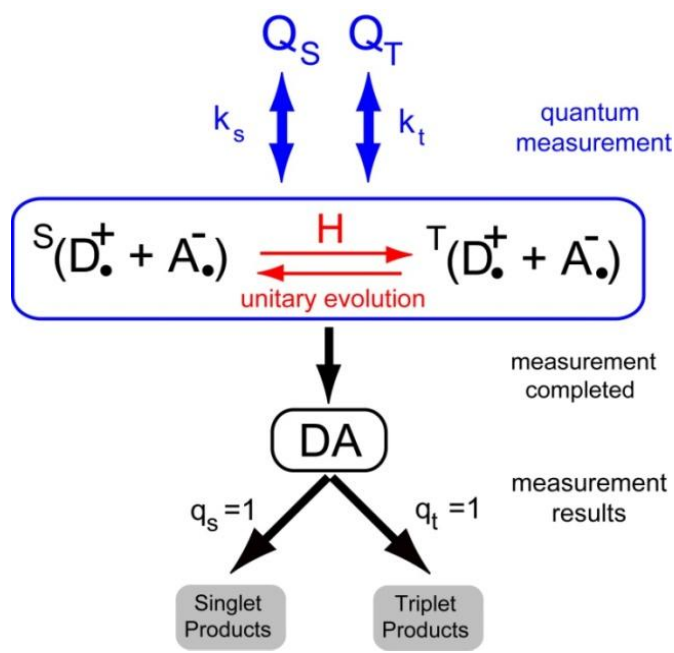

**Figure 1 Quantum and classical dynamics in the radical-ion-pair magnetoreception mechanism**. The radical-ion pair is formed by a donor and acceptor molecule, D and A, respectively. The singlet and triplet states of the pair evolve in a coherent superposition induced by the magnetic interaction Hamiltonian $H$, the so-called singlet-triplet (S-T) coherence. This is interrupted by the quantum measurement operators $Q_S$ and $Q_T$, which continuously interrogate the system at the respective measurement rates $k_s$ and $k_t$. When the measurement produces a definite result, i.e. an eigenstate of $Q_S$, the radical-ion pair can recombine into the corresponding singlet or triplet state products.

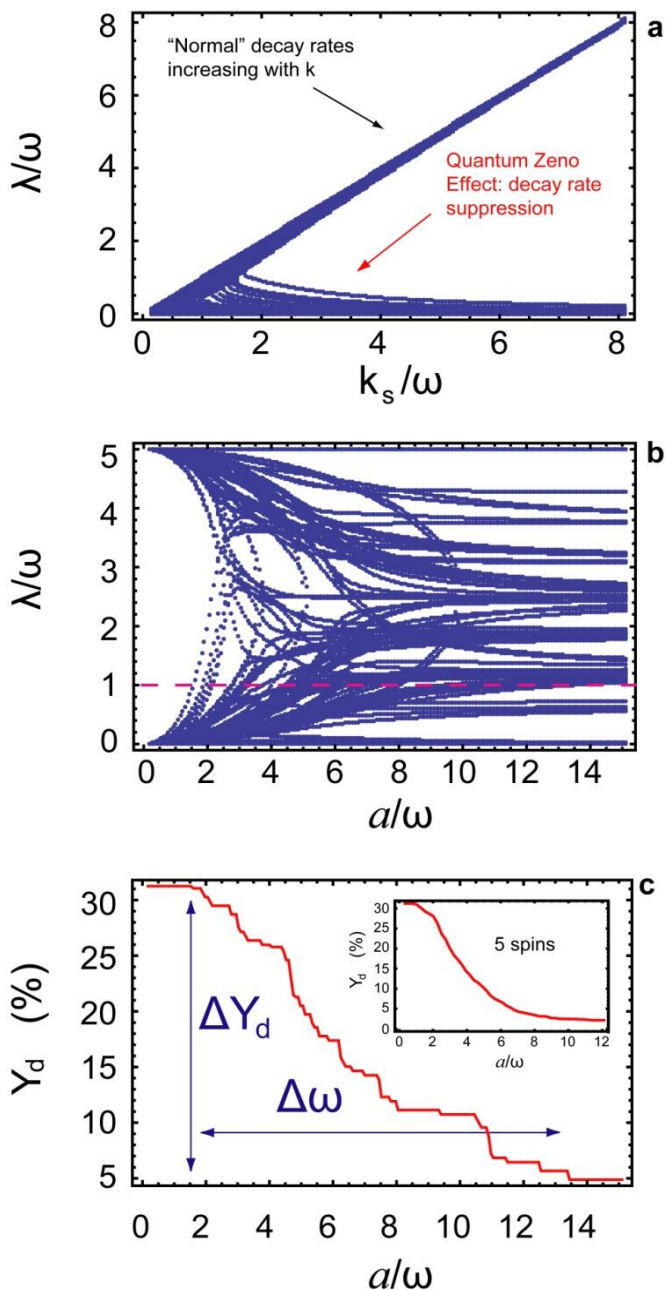

**Figure 2 Counting decay rates of the signlet-triplet coherence**. **(a)** Decay rates of the eigenvalue spectrum of the Liouville equation (4) normalized to the Larmor frequency $\omega$, as a function of the recombination rate $k_s$ (also normalized to $\omega$). About 60% of the 256 decay rates manifest the quantum Zeno effect, i.e. they decrease with increasing recombination rate. The rest follow the expected behavior, that is, they increase proportionally with the rate $k_s$. The calculation was performed for $\theta=0$ (magnetic field parallel to the molecule's z-axis) and a hyperfine frequency $a/\omega=1$. **(b)** For a particular value of $k_s$ well into the quantum Zeno regime (here $k_s/\omega=5$) we plot the decay rates as a function of the hyperfine to Larmor frequency ratio, $a/\omega$. The crossing of decay rates seen to occur is responsible for the physical realization of the sensitivity window. **(c)**

Half the percentage of the long-lived decay rates (those that lie below the dashed line in Fig.2b) are plotted versus the ratio $a/\omega$. This percentage represents the drop in the yield of singlet-state products, which is seen to change only in a window of Larmor frequencies, given a molecule-specific value of the hyperfine frequency. The discontinuous structure of the yield is due to the small number of eigenvalues ($4^4$=256) of the considered 4-spin system. In the inset, the dependence is seen to become much smoother just by adding one more nuclear spin.

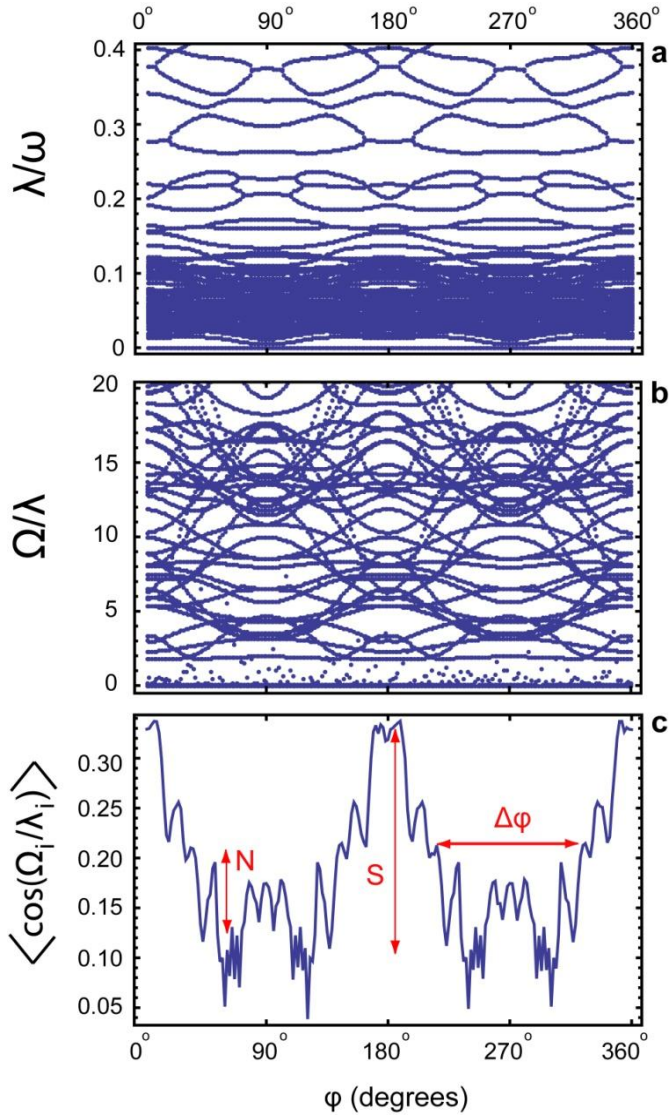

**Figure 3 Angular Dependence of Magnetoreception – Avian Heading Error**. **(a)** Long-lived decay rates (normalized to the Larmor frequency $\omega$) as a function of heading angle $\varphi$ (only a subset of the decay rates is shown for clarity). The heading angle is the angle between the magnetic field's projection on the RIP-fixed x-y plane and the x-axis. The RIP-fixed coordinate system is defined by the anisotropy of the hyperfine interactions. **(b)** This is a plot of the phases, i.e. the product of the precession frequency $\Omega$ with the mean survival time $1/\lambda$ for each eigenvalue. **(c)** The average over all long-lived eigenvalues of the cosine of the phases, as a function of $\varphi$. The noisy behavior limits the avian specie's ability to precisely determine heading, a fact corroborated by the measured angle dispersion in magnetic orientation experiments. The plot is reflection-symmetric about 180°, proving the experimental finding that avian species are insensitive to magnetic field polarity. The experimentally observed heading error of about 30° readily follows from the peak's signal-to-noise ratio and the angular width. The noise is due to the superposition of a large number of oscillating terms, and is found to persist in a system with 5 spins, which has 4 times more eigenvalues. The calculations were done for an inclination angle θ=45$^o$, where heading-angle sensitivity is maximum, $k/\omega$=5 and $a/\omega$=1.